\documentstyle[twoside,fleqn,jhuwkshp,psfig]{article}
\newcommand{\etal}{{\it et al.} }
\newcommand{\asca}{{\it ASCA} }
\newcommand{\ascap}{{\it ASCA}}

\newcommand{\bbxrt}{{\it BBXRT} }

\newcommand{\chandra}{{\it Chandra} }

\newcommand{\hetg}{{\it HETGS} }
\newcommand{\hetgp}{{\it HETGS}}
\newcommand{\letg}{{\it LETGS} }
\newcommand{\letgp}{{\it LETGS}}

\newcommand{\fekalfa}{{Fe~K$\alpha$} }

\newcommand{\fexxv}{Fe~{\sc xxv} }

\newcommand{\fexxvi}{Fe~{\sc xxvi} }
\newcommand{\fexxvip}{Fe~{\sc xxvi}}
\newcommand{\feklya}{{Fe~{\sc xxvi }~Ly$\alpha$} }

\newcommand{\feklyap}{{Fe~{\sc xxvi}~Ly$\alpha$}}
\newcommand{\resonetwo}{{$1s^{2}-1s2p$} }
\newcommand{\resonetwop}{{$1s^{2}-1s2p$}}

\newcommand{\figprefit}{{Fig.~1} }
\newcommand{\figprefitp}{{Fig.~1}}

\newcommand{\figbotharms}{{Fig.~2} }
\newcommand{\figbotharmsp}{{Fig.~2}}

\newcommand{\fighegmegcmpp}{{Fig.~3}}
\newcommand{\figdiskfit}{{Fig.~6} }

\newcommand{\figlinecontours}{{Fig.~5} }
\newcommand{\figlinecontoursp}{{Fig.~5}}
\newcommand{\figabscontours}{{Fig.~4} }
\newcommand{\figabscontoursp}{{Fig.~4}}

\newcommand{\figascalegratsp}{{Fig.~7}}

\newcommand{\figascalegprofilesp}{{Fig.~8}}
\newcommand{\figmultimissioncont}{{Fig.~9} }
\newcommand{\figmultimissioncontp}{{Fig.~9}}
\newcommand{\figascaivsi}{{Fig.~10} }
\newcommand{\figascaivsip}{{Fig.~10}}

\newcommand{\tablediskfits}{Table~2 }
\newcommand{\tablediskfitsp}{Table~2}
\newcommand{\tablegaussfits}{Table~1 }
\newcommand{\tablegaussfitsp}{Table~1}
\newcommand{\tablelegfits}{Table~3 }
\newcommand{\tablelegfitsp}{Table~3}
\newcommand{\tableascafits}{Table~4 }

\newcommand{\src}{E~1821$+$643 }
\newcommand{\srcp}{E~1821$+$643}

\begin{document}

\title{IRON K FEATURES IN THE
QUASAR E~1821$+$643: EVIDENCE FOR GRAVITATIONALLY REDSHIFTED ABSORPTION? 
}

\author{Tahir Yaqoob\address{\it Department of Physics and Astronomy,
Johns Hopkins University, Baltimore, MD 21218.}  
\address{\it Laboratory for High Energy Astrophysics,
NASA/Goddard Space Flight Center, Greenbelt, MD 20771.}
 and Peter Serlemitsos$^{\rm b}$ }


\begin{abstract}
\vspace{-3mm}
\centerline{\bf Abstract}

We report
a \chandra high-energy grating detection of a narrow, redshifted absorption line
superimposed on the red wing of a broad Fe K line in the $z=0.297$
quasar \srcp. The absorption line is detected at a confidence level,
estimated by two different methods, in the range
$\sim 2-3\sigma$. Although the detection significance is not
high enough to exclude a non-astrophysical origin,
accounting for the absorption feature when modeling the
X-ray spectrum implies that the Fe-K emission line is broad,
and consistent with an origin in a relativistic accretion disk.
Ignoring the apparent absorption feature leads to the conclusion
that the Fe-K emission line is narrower, and also affects
the inferred peak energy of the line (and hence the inferred ionization
state of Fe).
If the absorption line (at $\sim 6.2$~keV in the quasar
frame) is real, we argue
that it could be due to gravitationally redshifted
 \fexxv or \fexxvi resonance absorption
within $\sim 10-20$ gravitational radii of the putative central
black hole. The absorption line is not detected in earlier
\asca and \chandra low-energy grating observations, but the
absorption line is not unequivocally ruled out by these data.
The \chandra high-energy grating Fe-K emission line is consistent
with an
origin predominantly in Fe~{\sc i--xvii} or so. In an \asca
observation eight years earlier, the Fe-K line peaked at $\sim 6.6$~keV,
closer to the energies of He-like Fe triplet lines. Further, in a
\chandra low-energy grating observation the Fe-K line profile was
double-peaked, one peak corresponding to Fe~{\sc i--xvii} or so,
the other peak to \feklyap. Such a wide range in ionization state of
Fe is not ruled out by the HEG and \asca data either,
and is suggestive
of a complex structure for the line-emitter.


\centerline{\it To appear in the Astrophysical Journal, 20 April 2005}

\end{abstract}
\vspace{5mm}
\maketitle

\section{INTRODUCTION}\label{intro}

The high-luminosity ($L_{\rm 2-10 \ keV} \sim 3-4 \times 10^{45}
\rm \ ergs \ s^{-1}$) quasar \src ($z=0.297$) is one of the
highest redshift quasars to exhibit strong \fekalfa line emission.
The source has been observed by every X-ray astronomy mission since {\it EXOSAT},
but until recently, the contribution
to the \fekalfa line emission from the cluster that the quasar is
located in, was highly uncertain (e.g. see Saxton \etal 1997 and references therein).
Fang \etal (2002) finally
showed, from a \chandra observation,
that the cluster makes a negligible contribution
to line emission
at 6.4~keV (the energy of Fe~{\sc i}~$K\alpha$), and at most
$\sim 3\%$ at \feklya ($\sim 6.9$~keV).

In this paper we show that one interpretation of the \chandra data
for \src
is that the \fekalfa emission line is relativistically broadened,
with an absorption feature at $\sim 6.2$~keV (quasar frame)
superimposed on the broad emission line profile.
Whilst blueshifted absorption features are common in active galactic
nuclei (AGNs), a redshifted absorption feature has been reported
for only one case so far (NGC~3516, Nandra \etal 1999).
The absorption feature in \src could be due to
resonance absorption in highly ionized Fe, in the form of an inflow, matter
crossing the line-of-sight obliquely, or an outflow that is close
enough to the putative central black hole that the absorption
line suffers a strong gravitational redshift. The latter scenario
would of course have very important implications for the study of
the central engine with future missions.
Whatever the origin of the absorption line, its presence in the HEG
spectrum affects modeling of the Fe-K {\it emission} line
at $\sim 6.4$~keV and
therefore cannot be ignored.
The absorption
may be of a transitory nature since it is not detected in
another \chandra observation, nor was it detected with {\it ASCA}.
However, each of these latter data sets were not as sensitive to
the absorption line and we will show that they do not
unequivocally rule it out. Our analysis also shows a
significant detection of Fe-K line emission at $\sim 6.9$~keV
in \chandra low-energy grating (LEG) data (in addition to the line at $\sim 6.4$~keV).
Likely to be associated with \feklyap, this line is weak in
the HEG data and not detected in the \asca data, but both
data sets are statistically consistent with the
LEG data. The overall line profile in the \asca data has
a broad peak at $\sim 6.6$~keV but multiple
emission lines cannot be resolved.

The paper is organized as follows. In \S\ref{data} we describe the
\chandra high-energy grating data and its reduction; in \S\ref{spectralfitting}
we describe modeling of the Fe-K emission and absorption lines in these
data, including a detailed discussion of the statistical
significance of the absorption line. In \S\ref{otherdata} we
discuss the Fe-K emission and absorption features with respect to
two other important data sets for \srcp, namely from observations
with the \chandra LEG and with \ascap.
In \S\ref{absorigin} we discuss possibilities for the origin of the
absorption line, and finally, in \S\ref{conclusions} we present our
conclusions.

\section{THE \chandra HEG DATA}
\label{data}

\src was observed with \chandra starting 9 February, 2001,
from UT 14:28:55, for a duration of $\sim 100$~ks.
This \chandra observation was
made with the High-Energy Transmission Grating (or \hetg --
Markert, \etal 1995) in the focal plane of the
High Resolution Mirror Assembly. The \chandra \hetg affords
the best spectral resolution in the $\sim 6-7$~keV Fe-K band currently
available ($\sim 39$~eV, or $1860 \rm \ km \ s^{-1}$ FWHM
at 6.4 keV). \hetg consists of two grating assemblies,
a High-Energy Grating (HEG) and a Medium-Energy Grating (MEG),
and it is the HEG that achieves this spectral resolution.
The MEG spectral resolution
is only half as good as that of the HEG. The HEG also has a
higher effective area in the Fe-K band, so our study will
focus principally on the HEG data.
The HEG and MEG energy bands are 0.4--10~keV and 0.7--10~keV
respectively, but the effective area falls off rapidly with energy near both ends of each bandpass.
The \chandra data
were processed with version R4CU5UPD14.1
of the processing software, {\tt CALDB} version 2.2
was used, and the telescope responses made using
{\tt ciao 2.1.3}\footnote{http://asc.harvard.edu/ciao}.
Otherwise,
HEG and MEG lightcurves and spectra
were made exactly as described in Yaqoob \etal (2003a).
We used only the first orders of the grating data (combining
the positive and negative arms, unless otherwise stated). The zeroth order data
were piled up and the higher orders contain much fewer counts
than the first order. The mean count rates
(in the full energy band of each grating), over the entire
\chandra observation were $0.1346 \pm 0.0023$ ct/s and
$0.2938 \pm 0.0027$ ct/s for HEG and MEG respectively.
HEG and MEG spectra extracted over the entire observation,
resulted in a net exposure time of 99.620 ks.
Background was not subtracted since it is negligible
in the energy region of interest.
The source flux showed little variability over
the entire duration of the campaign. For example,
for HEG plus MEG lightcurves binned at 1024 s,
the excess variance above the
expectation for Poisson noise
(e.g. see Turner \etal 1999)  was $(-0.4 \pm 3.5) \times 10^{-4}$,
consistent with zero.

\begin{figure*}[bth]
\vspace{10pt}
\centerline{\psfig{file=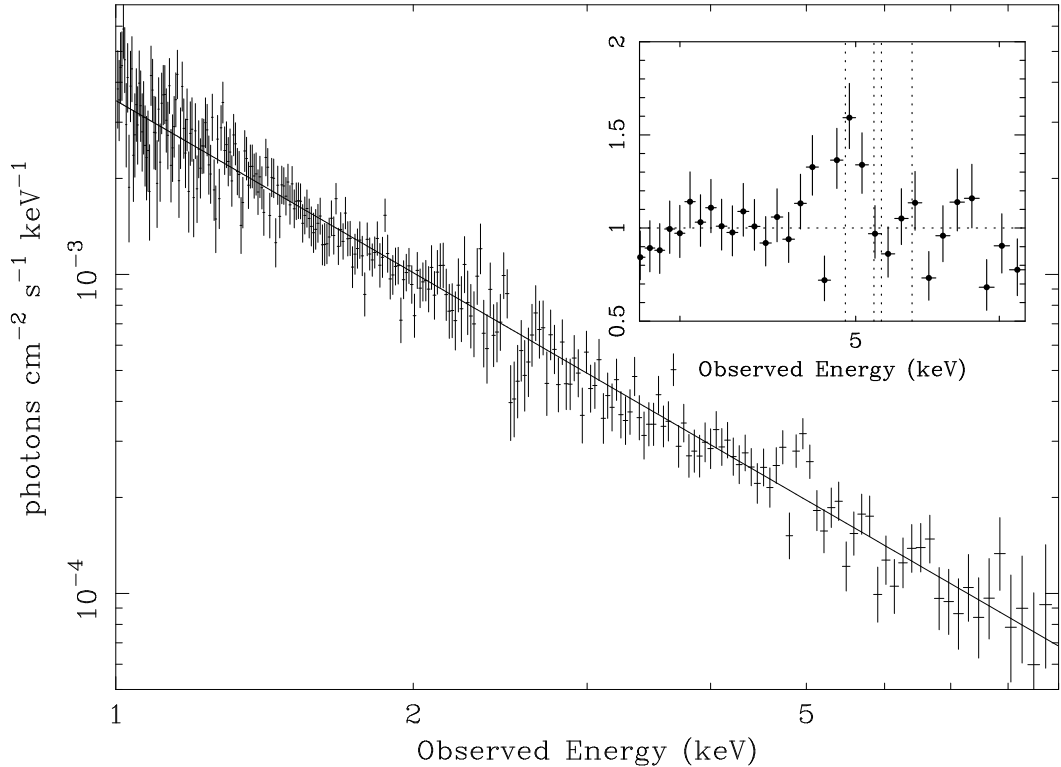,width=5.5in,height=4.0in}}
\caption{\footnotesize The HEG spectrum of \src (corrected for instrumental
response), binned at $0.04\AA$. The source redshift is 0.297.
The solid line is the best-fitting
power-law model fitted
over the
{\it observed} energy range 1.0--2.0, 2.5--9 keV (omitting the
2.0--2.5~keV region where the mirror response is not well modeled).
The energy range 1--9~keV corresponds to 1.3--11.7~keV in the
rest-frame.
The inset shows the ratio of the same spectrum to the
above power-law model in the Fe K region. An asymmetric, broad Fe-K line
is apparent, with a deep absorption feature superimposed on the profile.
The vertical dotted lines correspond (from left to right) to the
energies of the following transitions: Fe~{\sc i}~K$\alpha$, \fexxv
\resonetwo forbidden and
resonance lines, and \feklya (which correspond to
rest-frame energies of 6.400, 6.636, 6.700, and 6.966~keV
respectively).
}
\end{figure*}

\section{SPECTRAL FITTING OF THE BROAD Fe-K EMISSION LINE AND NARROW
ABSORPTION LINE}
\label{spectralfitting}

We used XSPEC v11.2 for spectral fitting (Arnaud 1996),
omitting the 2--2.5~keV
band due to systematic residuals around sharp changes
in the effective area of the telescope --
see Yaqoob \etal (2003a).
The response matrices
{\tt acisheg1D1999-07-22rmfN0004.fits} and
{\tt acismeg1D1999-07-22rmfN0004.fits}
(for the HEG and MEG respectively), combined with the telescope
response files
described above, were used to fold models through the instrument response
and thereby directly compare predicted and observed counts spectra.
Therefore, quoted fit parameters do {\it not} need to be corrected
for instrumental response.
For fitting purposes, the HEG and MEG spectra were binned at $0.01\AA$ and
$0.02\AA$ respectively. The bin size is comparable to the FWHM spectral
resolution ($0.012\AA$ and $0.023\AA$ for HEG and MEG respectively).
For clarity, spectral plots in this paper show larger bin sizes
than were used in the fitting (the bin size will be given case by case).

Since some energy bins may contain zero or few counts,
the $C$-statistic was used for minimization and,
unless otherwise stated, all statistical errors
correspond to 90\% confidence for one interesting parameter
($\Delta C = 2.706$). By definition, calculation of the
$C$-statistic requires only knowledge of the number of counts
in a bin, but for spectral plots, the error bars shown
correspond to asymmetric errors calculated using the approximations
of Gehrels (1986).
All model parameters will be
referred to the source frame using $z=0.297$,
unless otherwise stated.

We fitted the HEG spectrum in the 1--9~keV band, in the observed frame
(except for the 2--2.5~keV band).
Galactic absorption was not included
in the model fitting since it has a negligible effect above 1~keV.
First we fitted a simple power-law model.
This model is shown in \figprefit overlaid on HEG data that
have been corrected for the instrument resolution and binned at $0.04\AA$.
The inset in \figprefit shows the ratio of data to model.
It can be seen that there is broad,
asymmetric excess emission centered around $\sim 4.9$~keV ($\sim 6.4$~keV
quasar frame), likely corresponding
to a low-ionization state \fekalfa line.
A similar, but narrower
emission line has been reported by Fang \etal (2002) using the same data.
However, there also appears to be a deep absorption feature, up to about
$0.04\AA$ wide ($\sim 75$~eV at the observed energy), superimposed on the
red wing of the emission line, at $\sim 4.8$~keV in the observed
frame ($\sim 6.2$~keV in the quasar frame). The exact value of
the centroid energy of the feature, and associated statistical
errors, depend on the model of the emission line and
these matters are discussed
in \S\ref{gaussianfitting} and \S\ref{disklinefitting}.

\begin{figure}[tbh]
\vspace{10pt}
\centerline{\psfig{file=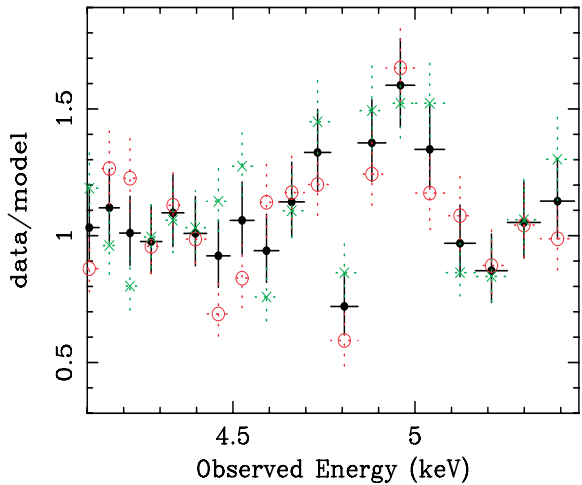,width=3.0in,height=2.5in}}
\caption{\footnotesize
Detection of the high-energy absorption line in the separate
$+1$ (green crosses) and $-1$ (red open circles) orders of the
\chandra HEG, compared
with the spectrum summed over both grating arms (black filled circles).
The absorption line is probably due to redshifted
He-like or H-like Fe absorption, equivalent to
an inflow velocity of $\sim 2 \times
10^{4} \ \rm km \ s^{-1}$
or $\sim 3 \times 10^{4} \ \rm km \ s^{-1}$
respectively (see text). Shown are the ratios of spectral
data (binned at $0.04\AA$) to the best-fitting power law
fitted to the spectrum combined from both $+1$ and $-1$ orders of the
grating (i.e. the same power-law model as shown in \figprefitp).
}
\end{figure}

\subsection{Reality of the Absorption Line}
\label{probcalcs}

We investigated whether the apparent absorption line could be due to
an instrumental artifact or
a statistical fluctuation.
We will describe this analysis in considerable detail since there
are important physical consequences if the absorption line is real.
Shown in \figbotharms are the data/model ratios for the
separate $+1$ and $-1$ orders, compared with the combined first-order
data/model ratio
(at bin sizes of $0.04\AA$), where the model is the simple
power law fitted to the combined first-order data, as
shown in \figprefit and described above.
\figbotharms shows that the absorption line is present, at some level, in both the
negative and positive arms of the grating.
Since the physical distance on the CCD detector of a photon event 
from the zeroth-order
image determines the energy of that photon event,
and the energy bin containing the
absorption line in the negative and positive arms have completely
different physical locations on the CCD array, an instrumental
artifact cannot explain the absorption line. Also, the effective
area of the HEG is smooth over the region the absorption line
(centered on $\sim 4.8$~keV, observed frame) is
detected: the combined $+1$ and $-1$ order effective area
varies by no more than 6\% over the observed energy range
4.6--5.0~keV. Yet the photon number in the $0.04\AA$ energy bin
in which the line is observed, deviates by at least 40\% compared
to adjacent bins. A mis-calibration of the effective
area would have to be present on extremely localized regions of
different CCDs, corresponding to precisely the same energy bin.
There are no other instrumental effects that we are aware of that
could explain the absorption feature either.

To quantify the possibility that a statistical fluctuation
occurred in both arms of the HEG, in the same energy bin, to conspire to yield an
apparent absorption line like that observed, we employed
two different methods to assess the statistical significance.
The question we are seeking an answer to is the same in
both cases. Namely, what is the probability of obtaining
a statistical
deviation in photon counts as large as that observed for the absorption line
in one arm of the grating {\it in any energy bin}
in the energy range searched, and {\it then} to
observe a deviation as large as that measured in the other arm of
the grating, {\it this time in the same energy bin}?

The first method we used is model-independent. For each first-order arm
of the HEG, we calculated the fractional deviation of the counts
in the $0.04\AA$ bin containing the absorption line, relative
to the mean counts in the two bins either side of it.
We must account for the total number of bins that were searched
that resulted in the detection of the absorption feature, given
that its energy does not correspond to a known atomic transition.
For this, we used a very conservative assumption, counting all
the energy bins that were used in spectral fitting.
So, in the 1--2~keV and 2.5--9~keV range, for the $+1$ order,
we calculated, for every bin,
the Poisson probability of obtaining the same fractional
deviation, say $X_{+}$ ($=58.1\%$),
as for the absorption line in that arm. For this,
we assumed the mean expected counts to be equal to the average of
the two bins surrounding the target bin. Then,
by combining these probabilities, we calculated the {\it total}
probability of observing the deviation, $X_{+}$, in
{\it any} of the energy bins in the energy range 1--2, 2.5--9~keV.
Next, we calculated the probability in the other HEG arm ($-1$ order)
of obtaining the observed deviation, $X_{-}$ ($=47.9\%$),
in the actual absorption-line bin,
relative to the average counts of the two surrounding bins. We obtained
a final probability of obtaining the actual deviations that were
observed, $X_{+}$ and $X_{-}$, in the $+1$ and $-1$
arms of the HEG respectively (and simultaneously), of $2.38\times 10^{-3}$
(i.e. this is the probability that {\it both} detections
were due to statistical fluctuations). This corresponds to an
overall significance of $3.04\sigma$ for detection of the absorption line.

We went further and estimated a significance using an even more conservative
assumption. This was based on the possibility that both bins surrounding the
absorption-line bin, in each HEG arm,
could both have been higher than the expectation values, due to statistical
fluctuations, thus artificially enhancing the depth of the absorption line.
Therefore, we repeated the above calculations after {\it reducing} the counts
in all bins by $1\sigma$, in both HEG arms, {\it except for the counts in the absorption
line bins}. Here, we used $1+\sqrt{(0.75+N)}$ for the $1\sigma$ error on the
counts, $N$ (see Gehrels 1986). Strictly speaking, this is not correct
for small $N$, but the bins either side of the absorption line in each arm
of the HEG have between 34 and 43 counts, so the effect of the
approximation on the final result is unimportant.
(Note that the MEG has between 17 and 29 counts per bin in the bins
either side of the absorption line, in each arm).
Using this more pessimistic
prescription we get $1.95\sigma$ for the significance of the absorption line,
which is consistent with the fact that we have effectively artificially
reduced the level of the reference `continuum' by $\sim 1\sigma$.

\begin{table}[hbt]
\begin{center}
\caption{Gaussian Line Fits to \chandra HEG Data 
for E~1821$+$643}
\end{center}
\label{tab:npagetab}
\begin{tabular}{@{}l@{\extracolsep{\fill}}r}
& \\
\hline
Parameter & Measurement \\
\hline
& \\
$C$-statistic & 1017.0 \\
Degrees of freedom & 968 \\
Emission Line Center Energy (keV) & $6.31^{+0.06}_{-0.05}$ \\
Emission Line Width, $\sigma$ (keV) & $0.18^{+0.05}_{-0.04}$ \\
Emission Line Velocity & $20145^{+5800}_{-4625}$ \\
(FWHM, $\rm km \ s^{-1}$) & \\
Emission Line Intensity & $5.2^{+1.3}_{-1.2}$ \\
($\rm 10^{-5} \ photons \ cm^{-2} \ s^{-1}$) & \\
Emission Line EW (eV) & $296^{+74}_{-69}$ \\
Absorption Line Center Energy & $6.228^{+0.011}_{-0.018}$ \\
(keV) & \\
Absorption Line Gaussian Width & $0.027^{+0.011}_{-0.007}$ \\
$\sigma$ ( keV) & \\
Absorption Line Velocity Width & $3040^{+1280}_{-780}$ \\
(FWHM, $\rm km \ s^{-1}$) & \\
Absorption Line Equivalent Width, & $54^{+12}_{-14}$ \\
EW (eV) & \\
Power-Law Photon Index $\Gamma$ & $1.82^{+0.03}_{-0.03}$ \\
& \\
\hline
& \\
\end{tabular}
{\small Simple power-law model plus a Gaussian emission line, and a Gaussian absorption line, fitted to the \chandra HEG data (see \S\ref{gaussianfitting} for details). All parameters (except 2--10~keV flux) are referred to the quasar frame ($z=0.297$).  Errors are 90\% confidence for one parameter ($\Delta C = 2.706$).  Velocities have been rounded to the nearest 5~$\rm km \ s^{-1}$.}
\end{table}

The second method we used to assess the significance of the line
employed Monte Carlo simulations.
We took a model for the continuum and emission line,
and folded it through the instrument response, so that we
could
measure deviations of the data relative to this model. We used the power-law
plus Gaussian emission line model described in \S\ref{gaussianfitting}
and \tablegaussfitsp, and set the absorption line
equivalent width (EW) in that
model to zero. Obviously the probabilities calculated from
the Monte Carlo simulations will depend somewhat on the
model but we will see that the results are
consistent with the model-independent method described above.

We measured negative deviations
relative to the continuum plus emission-line model of
20.5\% and 38.2\% in photon counts, in the $0.04\AA$ energy bin containing
the absorption line, in the $+1$ and $-1$ HEG orders respectively
and used these as inputs to the Monte Carlo calculations.
We ran $10^{6}$ simulations for every energy bin, for each HEG first-order
arm. We used the most conservative assumption again, for the number of
energy bins searched, namely the 1--2 and 2.5--9~keV range.
We measured the deviations relative to the model,
in each bin, and in each simulation, and thus
obtained a final probability that the observed absorption line
is due to chance statistical fluctuations, of $2.44 \times 10^{-2}$,
or a significance of $2.25\sigma$.
This is consistent with the estimate of 2--3~$\sigma$ from the
model-independent method described above.
We note that
these are all very conservative
estimates of the
significance, because in principle, we could have found the
absorption line in the real data by examining a smaller number of
energy bins.

From \figprefitp, it might appear as if there are deviations
as strong, or stronger, than those for the reported absorption line,
notably at $\sim 2.5$~keV, and above $\sim 6$~keV
(both observed frame).
We remind the reader that 2--2.5~keV is the region of greatest
changes in effective area, mainly due to the X-ray telescope (XRT),
so even small inaccuracies in the calibration of the effective
area can result in apparent emission or absorption features.
Indeed, one finds that the deviations at $\sim 2.5$~keV are not
consistent in the $+1$ and $-1$ HEG arms.
In fact, we already mentioned in \S\ref{spectralfitting} that we did not
use the 2--2.5~keV data in any of our analysis or simulations,
due to the difficult calibration of this region of the XRT response.

On the other hand, the deviations above $\sim 6$~keV in the
$+1$ and $-1$ HEG orders do
correlate with each other and may represent real structure. For example,
Fe-K edges are in the right energy range (6~keV corresponds to
$\sim 7$~keV in the quasar frame). However, the effective
area of the HEG (and MEG) drops quickly above 6~keV, and the
statistical significance
of these deviations is much less than that of the reported absorption
line, and is not sufficient to warrant further investigation.

\begin{figure}[tbh]
\vspace{10pt}
\centerline{\psfig{file=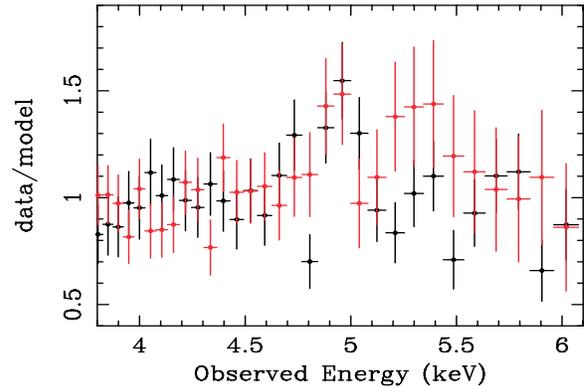,width=3.0in,height=2.0in}}
\caption{\footnotesize 
A comparison of the HEG data (black) with the MEG data (red)
in the Fe-K region in \srcp.
Shown are the ratios of spectral
data (binned at $0.04\AA$) to the best-fitting power-law
fitted to the HEG spectrum combined from both $+1$ and $-1$ orders of the
HEG (i.e. the same power-law model as shown in \figprefit and \figbotharmsp).
Although the absorption line does not appear to be
significantly detected in the MEG,
the probability of obtaining more counts than
expected in the absorption-line bin by chance (assuming the HEG
absorption-line equivalent width is the mean of a Poisson
distribution), is 46.4\%. Therefore the MEG data do not
rule out an absorption line.
}
\end{figure}

We also investigated whether the absorption line is
detected in the MEG data. A direct comparison of the MEG
and HEG data is given in \fighegmegcmpp. Shown are the ratios of MEG
and HEG data to a simple power-law continuum fitted in the
1.0-2.0, 2.5--9~keV bands. Although no absorption line is
evident in the MEG data, it is worth remembering that the
MEG effective area is a factor 1.5 less than that of the
HEG at the energy at which the HEG absorption line is
detected. We quantified the probability that the MEG
would {\it not} detect the absorption line, given
the HEG measurements.
The observed number of counts in the $0.04\AA$ bin
that would
contain the absorption line in the MEG spectrum
is 43. Taking the best-fitting HEG power-law plus Gaussian emission line,
and absorption-line model (see \S\ref{gaussianfitting}) and folding it
through the MEG instrumental response function,
and fitting for the overall cross-normalization difference
between HEG and MEG ($\sim 8\%$),  we estimated
the mean expected counts in the absorption-line bin to be 32.2.
The difference between expected and measured counts in that
bin is only $\sim 1.1\sigma$.
The Poisson probability of actually obtaining 33 or more
counts in that bin is then 46.6\%. In other words, this is the
probability that the MEG would observe an absorption line
that is apparently weaker than that measured by the HEG.
An alternative way of expressing the fact that the HEG and
MEG data are consistent with each other is that the
probability of measuring between 33--43 counts in the
absorption-line bin in the MEG is 44.1\%.

We conclude that the significance of the absorption line is not
low enough to
unequivocally attribute it to a statistical artifact, but neither is the
significance high enough to ascribe to an astrophysical origin
without caution. As we show below, the absorption line is significant
enough to affect modeling of the Fe-K {\it emission line}, so it
cannot be ignored.

\subsection{Gaussian Emission-Line Model}
\label{gaussianfitting}

Here we describe fitting the HEG data
(combined from the $+1$ and $-1$ order spectra) with
a power-law plus Gaussian emission line, with and without
an absorption line included.
If we do not include an absorption line, we obtain
Gaussian Fe-K emission-line parameters entirely consistent
with those measured by Fang \etal (2002). Namely, we obtained
a center energy of
$6.43^{+0.06}_{-0.05}$~keV, a Gaussian width of $0.10^{+0.07}_{-0.03}$ keV,
and an EW of $144^{+67}_{-57}$~eV (all parameters in the
quasar frame). Note that the EW given by Fang \etal (2002) is in
the observer's frame so must be multiplied by $(1+z)$ in order
to directly compare with ours.

On the other hand, if an inverted Gaussian is included to model
the absorption line, the emission line parameters are
different, because there is some apparent emission redward of the
absorption line, that is not modeled when the absorption line
is not included. This is because a single Gaussian model
worsens the fit if it is extended over the apparent absorption
in the data, so it remains narrow.
The power-law plus emission and absorption-line model involves
a total of eight free parameters, and their best-fitting values
and statistical errors are shown in \tablegaussfitsp.
It can be seen that, compared with an emission-line only
model, the peak energy of the emission line shifts
down by $\sim 150$~eV, and the intrinsic width is larger, because
when the an absorption line model is included,
there is excess emission on the red side of the absorption line
that is modeled as part of the emission line.
Also, since some of
the emission line is absorbed, a larger intrinsic intensity and EW are
required.

\begin{figure}[tbh]
\vspace{10pt}
\centerline{\psfig{file=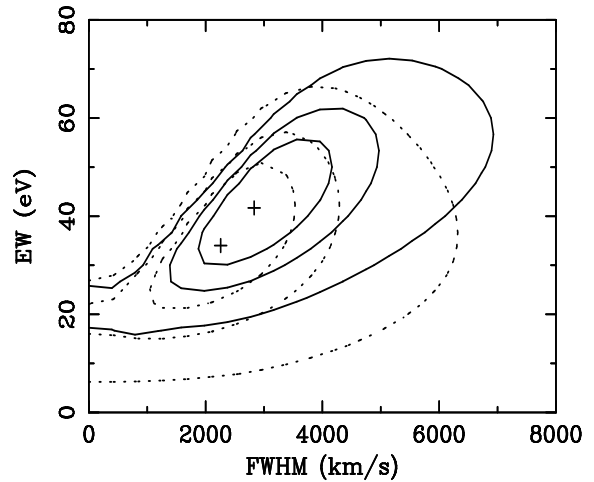,width=3.0in,height=2.5in}}
\caption{\footnotesize
Joint, two-parameter confidence contours (68\%, 90\%, and 99\%) for the
equivalent width (EW) and FWHM of the high-energy absorption-line feature in \src
 when the broad \fekalfa {\it emission line} is modeled in terms of a Gaussian (s
olid contours) and in terms of
relativistic disk line (dotted contours).
The absorption line is modeled with an inverted  Gaussian
in both cases.
In the quasar frame the center energy of the absorption line
is $6.228^{+0.011}_{-0.018}$~keV (Gaussian emission line)
or $6.220^{+0.018}_{-0.013}$~keV (disk emission line).
Details of all model parameters
are given in \tablegaussfits and \tablediskfitsp. The
contours were obtained from spectral fitting to the
HEG data (summed over $+1$ and $-1$ orders).
}
\end{figure}

\begin{figure}[tbh]
\vspace{10pt}
\centerline {\psfig{file=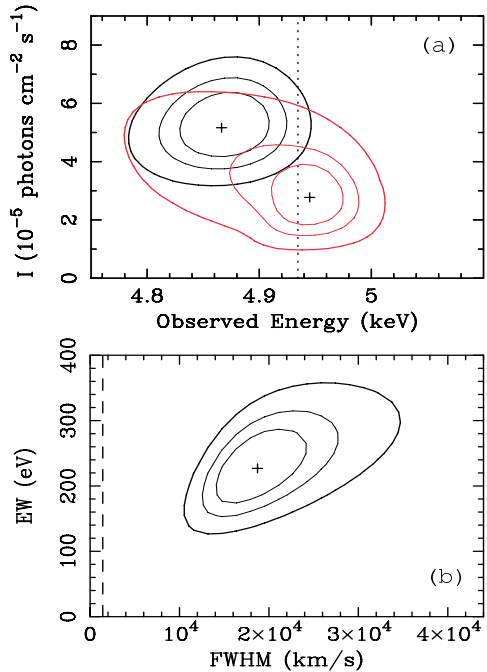,width=2.5in,height=3.5in}}
\caption{\footnotesize
(a) Contours corresponding to joint two-parameter
confidence levels of 68\%, 90\%, and 99\% for the intensity
and center energy (observed frame) of the broad \fekalfa line in \src when
it is modeled with a simple Gaussian.
Black contours correspond to the case
when an inverted Gaussian is included to model
the absorption line in the HEG data (see \figabscontoursp)
and red contours to the case when
the absorption line is omitted from the model and these
contours are entirely consistent with Fig.~4 of Fang \etal (2002).
{\it Caption is abridged: full caption given at the end of this article.}
}
\end{figure}

\figabscontours shows the joint two-parameter 68\%, 90\%, and 99\%
confidence level contours of
the absorption line EW versus its FWHM. The width depends somewhat
on how the {\it emission-line} is modeled (see \S\ref{disklinefitting}
for further discussion). For a Gaussian emission line model,
the FWHM is $\sim 3000 \rm \ km \ s^{-1}$
(see \tablegaussfitsp).
\figlinecontours (a) shows the joint two-parameter 68\%, 90\%, and 99\%
confidence level contours of
the Gaussian {\it emission-line} intensity versus observed center
energy, directly comparing the two cases,
with (black), and without (red) an absorption line
included in the model.
The (red) contours, for the model with no absorption line, are
consistent with those obtained by Fang \etal (2002), but note
that ours are plotted against {\it observed} energy in \figlinecontours (a),
but the same contours are shown against quasar-frame energy, in
\figmultimissioncontp.
\figlinecontours (b) shows
the EW of the emission line versus its FWHM
(when the absorption line is included in the model).
When modeled with a Gaussian, the emission line has a
FWHM of $\sim 20000  \rm \ km \ s^{-1}$ (see \tablegaussfitsp).

We note that Fang \etal (2002) reported the detection,
albeit marginal, of
an additional emission line at $6.94^{+0.05}_{-0.07}$~keV,
very likely due to \feklyap. Using combined HEG and MEG data
they obtained a $\Delta C$ of 9.2 upon the addition of
a three-parameter Gaussian to model the line, corresponding
to a confidence level of 97.3\%. Fang \etal (2002) measured a line
intensity and EW of $1.1 \pm 0.7 \times 10^{-5} \ \rm photons
 \ cm^{-2}
\ s^{-1}$ and $64 \pm 40$~eV respectively (we have corrected
the EW values of Fang \etal for cosmological redshift).
Only an upper limit was obtained for the intrinsic width,
0.15~keV.
Using the same model as Fang \etal (i.e.
a power law plus two Gaussians with no absorption line included),
we obtained consistent results for the high-energy iron line
from the HEG and MEG data.
Namely, a similar detection significance (98.0\%), a center
energy of $6.94 \pm 0.04$~keV, a line intensity of
$0.8^{+0.7}_{-0.5}  \times 10^{-5} \rm \ photons \ cm^{-2} \ s^{-1}$,
an EW of $40^{+35}_{-25}$~eV, and an intrinsic width
of $0.03^{+0.05}_{-0.02}$~keV. All statistical errors,
ours and those of Fang \etal (2002) quoted here for the high-energy
iron line are 90\% confidence for one interesting parameter.

\begin{table*}[t]
\caption{Relativistic Disk Line Fits to \chandra HEG Data for E~1821$+$643}
\label{tab:largetab}
\begin{tabular*}{\textwidth}{@{}l@{\extracolsep{\fill}}r}
& \\
\hline
Parameter & Measurement \\
\hline
& \\
$C$-statistic & 1014.5 \\
degrees of freedom & 966 \\
Disk Line Rest Energy (keV) & $6.57^{+0.01}_{-0.01}$ \\
                        & (6.51--6.68) \\
Disk Line Emissivity Index, $q$ & $2.69^{+0.19}_{-0.19}$ \\
                        & (2.36--3.08) \\
Outer Disk Radius, $R_{\rm out}$ & $>930$ \\
                        & ($>18$) \\
Disk Inclination, $\theta_{\rm obs}$ (degrees) & $0.0^{+0.4}_{-0.0}$ \\
                      & (0--27) \\
Disk Line Intensity ($\rm 10^{-5} \ photons \ cm^{-2} \ s^{-1}$) & $7.0^{+1.9}_{-
1.7}$ \\
                        & (3.6--10.2) \\
Disk Line EW (eV) & $209^{+51}_{-57}$ \\
                & (107--305) \\
Absorption Line Center Energy (keV) & $6.220^{+0.018}_{-0.013}$ \\
Absorption Line Gaussian Width, $\sigma$ (keV) & $0.021^{+0.012}_{-0.008}$ \\
Absorption Line Velocity Width, FWHM ($\rm km \ s^{-1}$) & $2385^{+1440}_{-950}$
\\
Absorption Line Equivalent Width, EW (eV) & $34^{+13}_{-13}$ \\
Power-Law Photon Index, $\Gamma$ & $1.84^{+0.03}_{-0.03}$ \\
2--10 keV Observed Flux ($10^{-11} \rm \ ergs \ cm^{-2} \ s^{-1}$) & 1.2 \\
2--10 keV Luminosity, Quasar Frame ($10^{45} \rm \ ergs \ s^{-1}$) & 3.3 \\
\hline
& \\
\end{tabular*}
{\small
Simple power-law model plus a relativistic disk
emission line, and a Gaussian absorption line,
fitted to the \chandra HEG data (see \S\ref{disklinefitting} for details).
All parameters (except 2--10~keV flux), refer to the quasar frame
($z=0.297$).
Errors are 90\% confidence for one parameter ($\Delta C = 2.706$);
90\% confidence, {\it five-parameter} ranges are also given for
the disk line parameters (i.e. $\Delta C = 9.24$) in parentheses.
Velocities have been rounded to the nearest 5~$\rm km \ s^{-1}$.
Intrinsic luminosity, in the 2--10~keV band in the quasar frame,
calculated using $H_{0} = 70 \rm \ km \ s^{-1} \ Mpc^{-1}$ and $\Lambda = 0.7$.}
\end{table*}
\subsection{Absorption Line Plus Relativistic Disk Line Model}
\label{disklinefitting}

Since the Fe-K line emission appears to be
asymmetric and peaks below 6.4~keV
($6.31^{+0.06}_{-0.05}$~keV, from \tablegaussfitsp),
we replaced the Gaussian emission line in the
model described above,
with a model of an \fekalfa emission line originating in a Keplerian disk
rotating around a Schwarzschild black hole (e.g. Fabian \etal 1989).
The inner radius of emission was fixed at $6r_{g}$
($r_{g} \equiv GM/c^{2}$) and the outer radius was a free parameter.
The line radial emissivity was assumed to be a power law
(line intensity proportional to $r^{-q}$), with the index, $q$, free.
In the disk rest-frame the emission-line energy was allowed
to vary between the \fekalfa line
energies for Fe~{\sc i} and \fexxvip, allowing a $\sim 10$~eV margin
on either side (i.e. 6.39--6.98~keV).
The disk inclination angle and overall line intensity were also free
parameters. Thus, there were a total of five free parameters
for the disk line model.

The best-fitting parameters of the absorption line,
disk emission line, and the continuum
are shown in \tablediskfitsp, and
the best-fitting model is shown in \figdiskfit (a). The counts spectrum
with the best-fitting model folded through the instrument response and
overlaid on the data is shown in \figdiskfit (b), whilst \figdiskfit (c)
shows the ratio of the data to the best-fitting model. Note that
since the $C$-statistic varied erratically
(for $|\Delta C| <5$) for some of the disk line
parameters, \tablediskfits also shows the 90\% confidence, five-parameter
ranges (i.e. $\Delta C = 9.24$).
The quasar-frame absorption-line parameters from the fit are
$6.220^{+0.018}_{-0.013}$~keV (center energy), $21^{+12}_{-8}$~eV
(Gaussian width), and $34 \pm 13$~eV (equivalent width, or EW).
\figabscontours shows joint two-parameter confidence contours
for the EW versus FWHM of the absorption line (dotted contours), directly
compared with the contours obtained when the emission line
was modeled with a Gaussian (\S\ref{gaussianfitting}).
The absorption-line parameters are a little different for the two cases,
(notably, the absorption-line EW is smaller when the emission
is modeled with a disk line).

\begin{figure}[tbh]
\vspace{10pt}
\centerline{\psfig{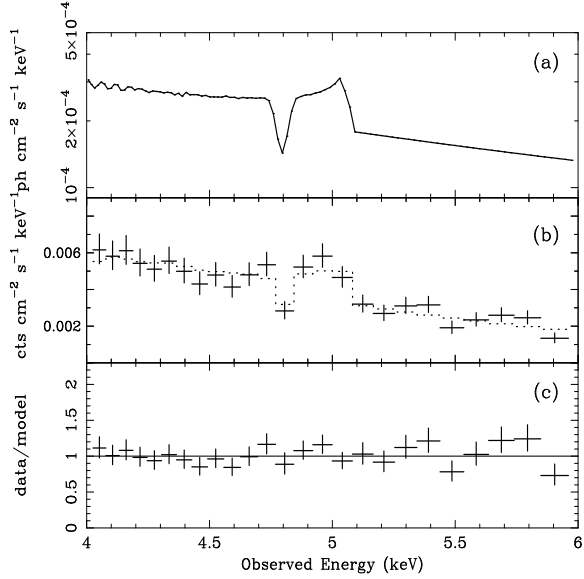}}
\caption{\footnotesize
(a) Best-fitting power-law plus disk emission-line model for
the HEG data for \srcp,
including an inverted Gaussian for the absorption line.
(b) HEG counts spectrum
(summed over $+1$ and $-1$ orders),
with the best-fitting model (solid line) in (a)
overlaid on the data, after folding through the instrument response.
(c) Ratio of the spectral data to the best-fitting model in (b).
In both (b) and (c) the spectra are binned at $0.04\AA$.
}
\end{figure}

We compare our
best-fitting disk line parameters to those of Fang \etal (2002).
Our inclination angle (zero degrees) appears to be smaller,
but since $C$ varies erratically around this best-fitting angle,
when one considers the 90\%, five-parameter upper limit (27 degrees),
our result is consistent with that of Fang \etal (2002), who obtained
$21.5 ^{+6.9}_{-10.4}$ degrees. Our EW of $209^{+51}_{-57}$~eV is
consistent with that obtained by Fang \etal (2002),
who obtained $168^{+59}_{-72}$~eV, which has to be
be multiplied by $(1+z)$ in order to
correct for cosmological redshift.
We do, however, require a
larger line flux in order to compensate for the absorption line
included in the model. We fitted the disk-line model without
the absorption line and obtained a line intensity of
$4.3^{+1.4}_{-1.4} \times 10^{-5} \ \rm \ photons \ cm^{-2} \ s^{-1}$,
consistent with the Fang \etal (2002) value of
$3.2^{+1.1}_{-1.4} \times 10^{-5} \ \rm \ photons \ cm^{-2} \ s^{-1}$
(again, Fang \etal did not correct for cosmological redshift
so their value has to be increased by 1.297 before comparing with
ours).

Although the signal-to-noise of the HEG data is limited,
the data are quite sensitive to the disk inclination
angle not being too large and the disk outer radius not being
too small. If the inclination angle is much greater than
$\sim 30^{\circ}$, and/or the outer radius much less than
$\sim 20r_{g}$, then the disk line would be too broad to fit the data.
We note that the inclusion of a Compton-reflection continuum
to the model described above has a negligible effect on the disk line
parameters because the additional component is not required
($|\Delta C| < 1$) due to poor statistics at the high-energy end of
the spectrum. For solar abundances, we obtain a
90\% confidence, one-parameter upper limit
on the solid angle of the reflector of $0.7(2\pi)$. For
an Fe abundance of three times solar, this upper limit is $0.95 (2\pi)$.
Ionization of the disk and/or inclusion of relativistic smearing effects
increases this upper limit but still has no effect
on the line parameters because structure in the continuum produced
by Compton reflection that could potentially
affect the line parameters, is not required by the data.

\section{Fe-K EMISSION \& ABSORPTION IN OTHER DATA SETS}
\label{otherdata}

There are two other principal data sets
for \src
that we can examine in order to compare with the
\chandra \hetg data, namely data from observations
with the \chandra low-energy grating spectrometer ({\it LETGS})
and {\it ASCA}. Except for a \bbxrt observation, that
was very short, and in which the Fe-K emission line was
barely detected
(Yaqoob \etal 1993), all observations of
\src prior to the {\it ASCA} observation
were made with proportional counters, which had poor spectral resolution.
\src has not been observed by {\it XMM-Newton}
(although it is present, but too far off-axis, in an observation pointed
at a nearby brown dwarf).
Below we describe results from observations with the \chandra {\it LETGS}
and \asca in turn.

\subsection{Chandra LEG Observation}
\label{legfitting}

\src was observed by the \chandra
low-energy grating spectrometer ({\it LETGS}) in 2001, January 17--24,
resulting in a deep exposure of this object.
Although the low-energy grating (LEG) has a spectral resolution
of only $0.05\AA$ FWHM (or $\sim 150$~eV at 6~keV),
it has an effective area a factor of $\sim 2$
larger than that of the HEG at the observed energy of the line.
The ACIS CCDs were used as the
detector for the gratings in this observation.
The low-energy spectrum from this observation
has been discussed by Mathur \etal (2003), but details of the Fe-K complex
from this data set have not been discussed in the literature.
We obtained a net exposure time of 470.8~ks, and the source
showed negligible flux variability during the entire observation.

\begin{figure}[tbh]
\vspace{10pt}
\centerline{\psfig{file=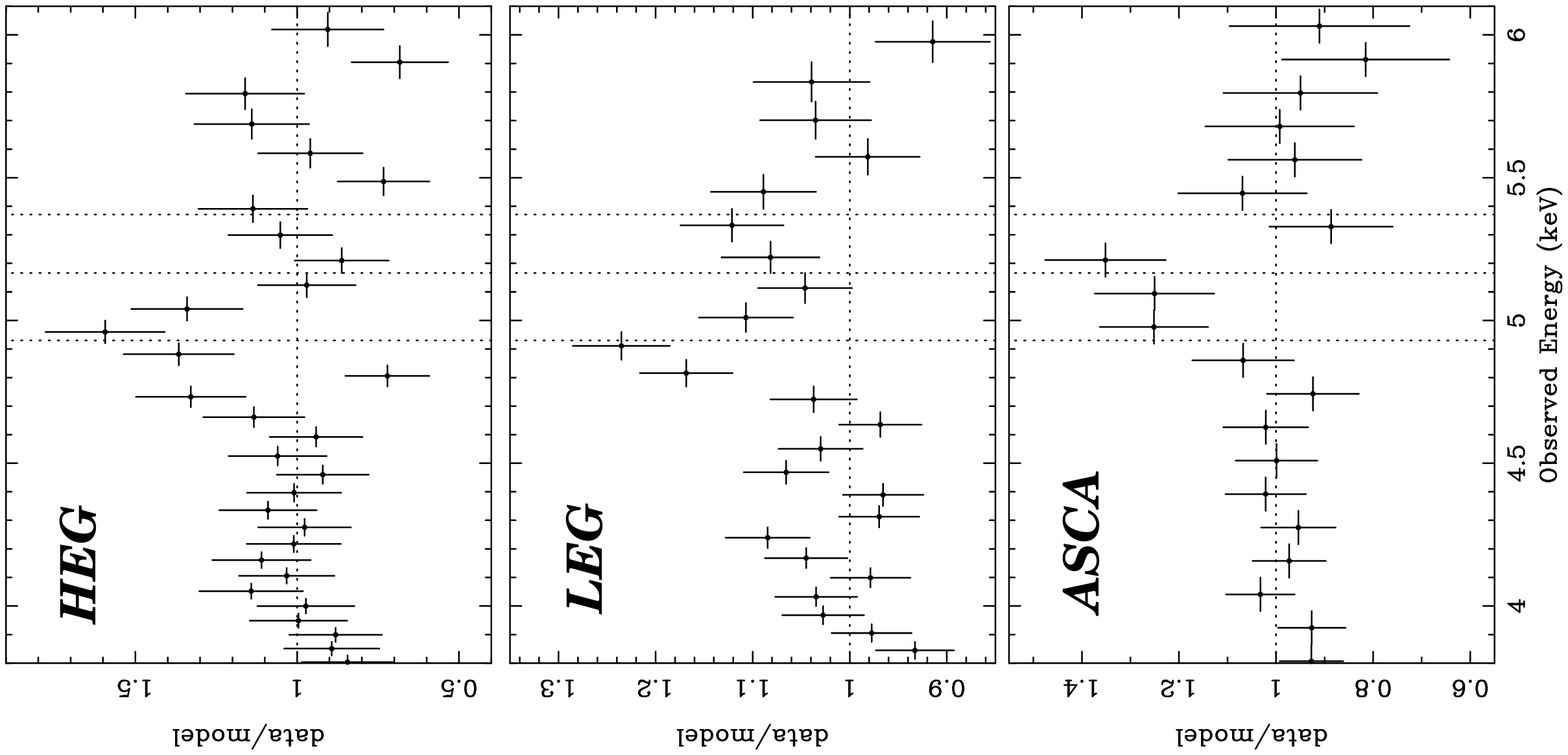,width=3.0in,height=4.5in,angle=270}}
\caption{\footnotesize
Comparison of the Fe-K complex in \src between non-contemporaneous
\chandra HEG, \chandra LEG, and \asca (SIS) data (see \S\ref{otherdata}).
Shown are the ratios of data to a simple power-law model.
{\it Caption is abridged: full caption given at the end of this article.}
}
\end{figure}

The ratio of the LEG data (around the Fe-K region),
to a single power law fitted over the
1--2, 2.5--9~keV observed energy interval, is shown in \figascalegratsp.
An estimate of the line profile itself, with the above
continuum subtracted off is shown in \figascalegprofilesp.
The latter plot also shows the LEG data directly compared with the
HEG data, showing that the overall line flux in the HEG is
somewhat larger.
It can be seen that there are {\it two} apparent emission peaks
in the LEG data, not one.
One peak is centered near 6.4~keV (quasar frame)  and the
other near the energy of the \feklya transition,
which is at 6.966~keV (rest energy).
The second emission-line peak in the LEG data occurs across
several resolution elements and is present in both the
$+1$ and $-1$ orders of the grating data. It also has a
high statistical significance: the difference in the
fit statistic between a single and double-Gaussian model
for the line emission is 18.3 (for two additional free
parameters in the double-Gaussian model).

\begin{figure}[tbh]
\vspace{-4mm}
\centerline{\psfig{file=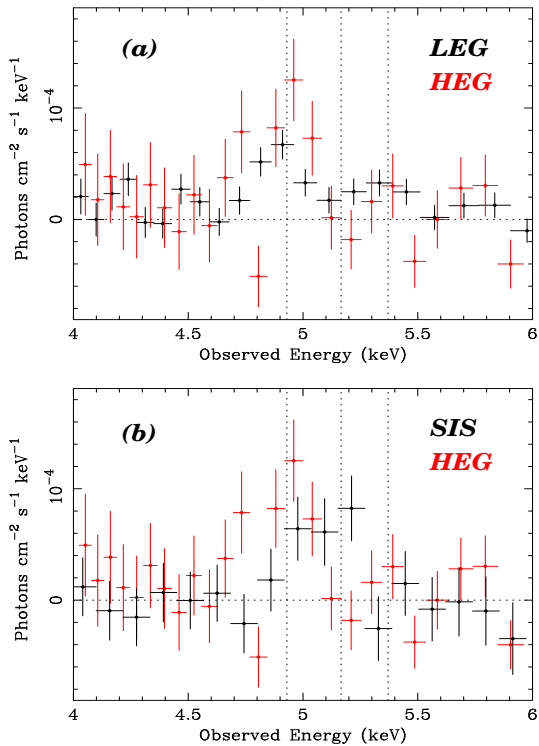,width=3.8in,height=4.5in,angle=180.0}}
\vspace{-15mm}
\caption{\footnotesize
Comparison of the Fe-K complex in \src between non-contemporaneous
\chandra HEG, \chandra LEG, and \asca (SIS) observations
(see \S\ref{otherdata}).
Shown are the {\it absolute} fluxes after the best-fitting power-law
continua (as fitted in \figascalegratsp) are subtracted.
Therefore, these plots can be used
to assess absolute intensity changes in the line emission.
(a) \chandra LEG data (black) compared directly with the \chandra
HEG data (red). (b) \asca SIS data (black) compared directly with the \chandra
HEG data (red).
}
\end{figure}

\tablelegfits shows the best-fitting
parameters and their statistical errors, obtained from
fitting a power-law
plus double-Gaussian model to the LEG data.
It can be seen that the energies of the emission-line peaks
are very well constrained.
The fact that the first and second emission-line peaks in
the LEG data are consistent with the energies of K lines
from Fe~{\sc i} and \fexxvi respectively, suggests
that they are two separate emission lines, rather than a single,
broad, disk line profile.

It would appear that the Fe-K emission-line profile
measured from the LEG is different
to that measured from the HEG data.
However, as pointed out in \S\ref{gaussianfitting},  Fang \etal (2002)
reported marginal evidence for a peak near the energy of \feklya in
the \chandra HEG data, which we confirmed
(\S\ref{gaussianfitting}) and which can be seen in our \chandra
HEG spectrum (e.g. \figprefitp). In fact, formally, the
intensity of the high-energy iron emission line measured by the
HEG is statistically consistent with that measured by the LEG
(compare HEG values in \S\ref{gaussianfitting} with
LEG values in \tablelegfitsp).

No absorption line at the position seen in the HEG data is
evident in the LEG data. However we can obtain an upper limit on
the EW of an absorption line if it were present, and had the
center energy and width of the line observed in the HEG (see \tablegaussfitsp).
We added an inverted Gaussian with the HEG parameters
to the double-Gaussian emission-line
model and obtained an upper limit on the EW of the absorption line
of 23~eV, to be compared with the HEG values of $54^{+12}_{-14}$
and $34^{+13}_{-13}$~eV, corresponding to Gaussian and disk-line
models of the Fe-K {\it emission}, respectively
(all values are in the quasar frame). Thus, the LEG data do
not rule out the presence of the absorption line if the disk-line
model is more appropriate for the emission line than the Gaussian model.
Moreover,
strictly speaking, a {\it one-parameter} 90\% upper limit for
the LEG absorption-line EW is not appropriate since there are
a total of eight free parameters in the model (plus a continuum
normalization). The 90\%, eight-parameter upper limit on the LEG
included in the HEG model, then the 99\% HEG and LEG contours do
not overlap. Therefore, it is possible that the lower-energy peak
in the Fe-K line profile varied between the HEG and LEG observations.

\subsection{ASCA Observation}
\label{ascafitting}

\src was observed by \asca in 1993, June 19 and the results
of this observation have been discussed by Yamashita \etal (1997).
We reanalyzed the data, in a manner similar to that described
in Weaver, Gelbord, \& Yaqoob (2001). We restricted our analysis
to the two CCD Solid State Imaging Spectrometers (SIS)
since they had much better spectral resolution than the
Gas Imaging Spectrometers (GIS) aboard {\it ASCA}.
At 6~keV, the spectral resolution of the SIS was $\sim 150$~eV FWHM
at the time of the observation.
We combined data from the two SIS detectors and obtained a
net exposure time of 83.4~ks.
We used $\chi^{2}$ statistics for fitting the data since
the counts per bin are much higher than in the \chandra
grating spectra, and background has to be subtracted for
\asca data.

The ratio of the SIS data to a single power law fitted over the
1--9~keV observed energy interval is shown in \figascalegratsp.
This plot also shows the \asca data directly compared with the
HEG data.
An estimate of the line profile itself, with the above
continuum subtracted off is shown in \figascalegprofilesp.
The latter plot also shows the \asca data directly compared with the
HEG data, showing that the line appears to peak at a different
energy in each observation.
It can be seen that there is a broad emission peak in the \asca data,
centered near the He-like Fe triplet transitions
($\sim 6.6-6.7$~keV in the quasar frame).

\begin{table*}[t]
\caption{Double Gaussian Line Fits to \chandra LEG Data for E~1821$+$643}
\label{tab:largetab}
\begin{tabular*}{\textwidth}{@{}l@{\extracolsep{\fill}}r}
& \\
\hline
Parameter & Measurement \\
\hline
& \\
$C$-statistic & 256.2 \\
Degrees of freedom & 187 \\
$E_{1}$, First Emission Line Rest Energy (keV) & $6.35^{+0.03}_{-0.05}$ \\
$I_{1}$,  First Line Intensity ($\rm 10^{-5} \ photons \ cm^{-2} \ s^{-1}$) & $1.40^{+0.35}_{-0.30}$ \\
$EW_{1}$,  First Emission Line Equivalent Width (eV) & $89^{+23}_{-19}$ \\
$E_{2}$, Second Emission Line Rest Energy (keV) & $6.91^{+0.09}_{-0.09}$ \\
$I_{2}$,  Second Emission Line Intensity ($\rm 10^{-5} \ photons \ cm^{-2} \ s^{-1}$) & $0.82^{+0.31}_{-0.34}$ \\
$EW_{2}$, Second Emission Line Equivalent Width (eV) & $61^{+23}_{-26}$ \\
$\sigma$, Gaussian Emission Line Widths (keV) & $0.12^{+0.06}_{-0.03}$ \\
FWHM of Emission Lines ($\rm km \ s^{-1}$) & $13350^{+5050}_{-4140}$ \\
Absorption Line Equivalent Width, EW (eV) & $0^{+23}_{-0}$ \\
Power-Law Photon Index, $\Gamma$ & $1.87^{+0.01}_{-0.01}$ \\
2--10 keV Observed Flux ($10^{-11} \rm \ ergs \ cm^{-2} \ s^{-1}$) & 1.0 \\
2--10 keV Luminosity, Quasar Frame ($10^{45} \rm \ ergs \ s^{-1}$) & 2.7 \\
\hline
& \\
\end{tabular*}
{\small
Simple power-law model plus two Gaussian emission lines
fitted to the {\it Chandra} low-energy grating (LEG) data
(see \figascalegratsp, \figascalegprofilesp, \figmultimissioncontp,
and \S\ref{legfitting} for details).
All parameters (except 2--10~keV flux) refer to the quasar frame
($z=0.297$).
Errors are 90\% confidence for one parameter ($\Delta C = 2.706$).
The emission-line velocity widths (forced to be the same for
the two lines) were frozen for deriving the error ranges on the
other parameters, otherwise the fits became unstable.
The EW of a Gaussian absorption line added to the above model
is also given, where the center energy and width of the
absorption line are fixed at the respective values obtained from
the HEG data (see \tablegaussfitsp).
Velocities have been rounded to the nearest 5~$\rm km \ s^{-1}$.
Intrinsic luminosity, in the 2--10~keV band in the quasar frame,
was calculated using $H_{0} = 70 \rm \ km \ s^{-1} \ Mpc^{-1}$ and $\Lambda = 0.7$.}
\end{table*}

\tableascafits shows the results of fitting the \asca data
with a power-law plus Gaussian emission-line model.
The center energy of the line is $6.62^{+0.12}_{-0.12}$~keV,
and the EW is $140^{+78}_{-86}$~eV (both in the quasar frame). The
fitted to the {\it Chandra} low-energy grating (LEG) data
(see \figascalegratsp, \figascalegprofilesp, \figmultimissioncontp,
and \S\ref{legfitting} for details).
All parameters (except 2--10~keV flux) refer to the quasar frame
($z=0.297$).
Errors are 90\% confidence for one parameter ($\Delta C = 2.706$).
The emission-line velocity widths (forced to be the same for
the two lines) were frozen for deriving the error ranges on the
other parameters, otherwise the fits became unstable.
The EW of a Gaussian absorption line added to the above model
is also given, where the center energy and width of the
absorption line are fixed at the respective values obtained from
the HEG data (see \tablegaussfitsp).
Velocities have been rounded to the nearest 5~$\rm km \ s^{-1}$.
Intrinsic luminosity, in the 2--10~keV band in the quasar frame,
was calculated using $H_{0} = 70 \rm \ km \ s^{-1} \ Mpc^{-1}$ and $\Lambda = 0.7$.

\tableascafits shows the results of fitting the \asca data
with a power-law plus Gaussian emission-line model.
The center energy of the line is $6.62^{+0.12}_{-0.12}$~keV,
and the EW is $140^{+78}_{-86}$~eV (both in the quasar frame). The
line is unresolved with {\it ASCA}. Our results
are consistent with those of Yamashita \etal (1997), who
obtained a center energy of $6.58 \pm 0.05$~keV and an EW of
$100\pm50$~eV (both in the quasar frame).
\figmultimissioncont shows the 68\%, 90\%, and 99\% contours
of line intensity versus line energy obtained from
Gaussian-fitting of the emission line in the \asca data,
compared with confidence contours obtained from the HEG and LEG
data.
It can be seen that the 99\% confidence
contours obtained from modeling the principal line peak in the
HEG and \asca data
overlap (with or without an absorption line included in the
HEG model).
Therefore, at 99\% confidence, we cannot exclude the case that
the HEG and \asca emission-line profiles are consistent with
each other. At 99\% confidence, the LEG contours for the
peak at $\sim 6.4$~keV also overlap with the \asca
contours. Although the emission line at $\sim 6.9$~keV is
not detected in the \asca data, the presence of such a line
with the same energy and width as the in the LEG data
is not ruled out by the \asca data. Adding such an additional
Gaussian emission line to the \asca data,
even the 90\%, one-parameter upper limit on the intensity,
$1.3 \times 10^{-5} \ \rm photons \ cm^{-2} \ s^{-1}$,
is consistent with the LEG intensity
($0.8^{+0.3}_{-0.3} \times 10^{-5} \ \rm photons \ cm^{-2} \ s^{-1}$;
see \tablelegfitsp).

In order to test the \asca data for the
absorption line found in the HEG data,
we added an inverted Gaussian to our power-law plus single Gaussian
emission-line model in order to obtain an upper limit on the
EW of an absorption line with energy and width fixed at the
values obtained from the HEG data (see \tablegaussfitsp).
There were now a total five free parameters, plus a continuum
normalization.
The 90\%, one-parameter and five-parameter upper limits on the EW
are 82~eV and 138~eV (quasar frame) respectively.
Thus, the \asca data do not rule out the presence of an absorption
line that has the parameters that were measured by the \chandra HEG.

We also modeled the emission line in the \asca data with a
relativistic disk-line model, as was fitted to the HEG data
in \S\ref{disklinefitting}. Since the spectral resolution and
signal-to-noise of the data are limited, we fixed $r_{\rm in}=6r_{g}$,
$r_{\rm out}=1000r_{g}$, and $q=2$. The line energy in the disk
frame, its intensity, and the disk inclination angle were free
parameters. We obtained a best-fitting
line energy of $6.71^{+0.10}_{-0.16}$~keV (in the disk frame),
and an EW of $161^{+89}_{-67}$~eV (in the quasar frame), but
obtained only an upper limit on the inclination angle, of $30^{\circ}$
(90\%, one-parameter). The inclination angle upper limit is
consistent with the inclination angle inferred from disk-line
fits to the HEG data (\S\ref{disklinefitting}). However, the \asca data
strongly constrain the line energy in the disk frame to be near
that expected for He-like Fe triplet emission
(between $\sim 6.6$--6.7~keV). The
HEG data constrained the line energy to values appropriate for
$K\alpha$ transitions from much lower ionization states of Fe
(i.e. Fe~{\sc i}--Fe~{\sc xvii} or so).
This is not surprising since the HEG data are strongly peaked
near 6.4~keV, whilst the \asca data are strongly peaked near 6.6~keV
(quasar-frame quantities).

With the above \asca disk-line model, we obtained a somewhat smaller
upper limit on the EW of an absorption line with the parameters
measured from the HEG data, namely $53$~eV (90\% confidence, one-parameter).

\begin{figure}[tbh]
\vspace{10pt}
\centerline{\psfig{file=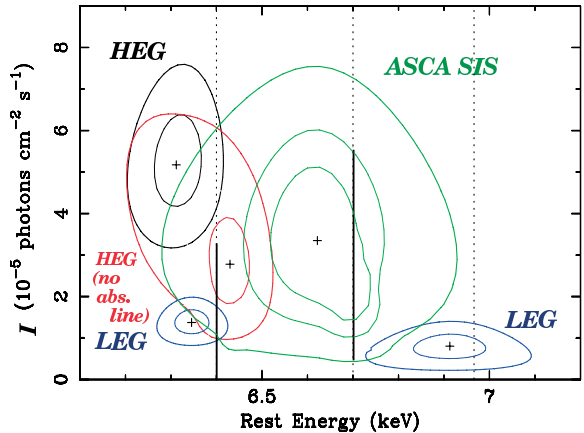,width=3.0in,height=2.25in}}
\caption{\footnotesize
Direct comparison of
joint, two-parameter, 68\%, and 99\% confidence contours of emission-line
intensity, $I$, versus line energy between
non-contemporaneous
\chandra HEG, \chandra LEG, and \asca (SIS) data (see \S\ref{otherdata}).
For \ascap, 90\% confidence contours are also shown.
In each case the data were modeled with a simple power law and
a Gaussian emission line (two in the case of the LEG in order to
model the two peaks - see \figascalegratsp).
The black HEG contours were obtained when an inverted gaussian
was included to model the absorption feature (the same contours
are shown in \figlinecontoursp). The red HEG contours were
obtained with no absorption feature included (as were the LEG and \asca contours). 
The \asca emission line can also be modeled as two discrete,
unresolved lines at 6.4~keV and 6.7~keV, and the 99\% two-parameter ranges
on the line intensities for this scenario are shown as solid vertical bars
at those energies (also see \figascaivsip).
{\it Caption is abridged: full caption given at the end of this article.}
}
\end{figure}

\begin{figure}[bht]
\vspace{10pt}
\centerline{\psfig{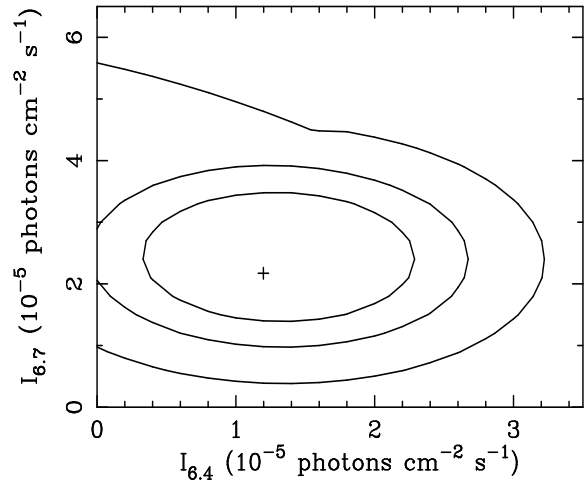}}
\caption{\footnotesize
The \asca emission line in \src can be modeled as two discrete,
unresolved lines at 6.4~keV and 6.7~keV, corresponding to
Fe~{\sc i}~K$\alpha$, \fexxv \resonetwo
resonance transitions respectively (see \S\ref{ascafitting} for
details). Shown are the
joint, two-parameter, 68\%, 90\%, and 99\% confidence contours of
the intensity of the 6.7~keV line ($I_{6.7}$) versus the
intensity of the 6.4~keV line ($I_{6.4}$). It can seen
that the 6.4~keV line is not required at $>90\%$ confidence but
is allowed to have an intensity high enough to be present
at the level detected in either the
HEG or LEG data (see \figmultimissioncontp).
}
\end{figure}

An alternative description of the \asca data is that
there are two, narrow, unresolved emission lines. We
fitted a dual-Gaussian model for the line emission, with
the line widths free but forced to be the same value for the two
lines. We fixed the energy of one line at 6.40~keV (Fe~{\sc i}~$K\alpha$),
and that of the other line at 6.70~keV (\fexxv resonance, \resonetwop).
We obtained equivalent widths of $42^{+37}_{-33}$~eV, and
$87^{+54}_{-34}$~eV respectively (quasar frame). The lower and upper bounds
on the line intensities
from this model are shown in \figmultimissioncontp, along with
intensity versus energy contours for the Gaussian models fitted
to HEG, LEG, and {\it ASCA} data, as described in previous sections.
\figascaivsi shows the 68\%, 90\%, and 99\% joint confidence
contours of the intensity of one line versus the other from the
dual-Gaussian model for two unresolved lines in the \asca data.
\figascaivsi shows that, in this model, the 6.7~keV line dominates the line profile
since the 6.4~keV is not actually required at 99\% confidence and higher.
We obtained an upper limit of 8,400 $\rm km \ s^{-1}$ FWHM on the width of
the lines in the dual-Gaussian model.

\section{ORIGIN OF THE ABSORPTION LINE}
\label{absorigin}

If the absorption line at $\sim 6.2$~keV detected in the HEG
data is real, variability is a possible
explanation of non-detection of the absorption line in the \asca
and LEG data (but we recall that neither of these data sets unequivocally
rule out the absorption line).
If the absorption line is real,
it could be due
to a redshifted resonance transition in highly ionized Fe.
The high ionization state would be consistent with the fact
that we find no significant soft X-ray absorption features in the HEG data
(see also Fang \etal 2002), and Mathur \etal (2003) came to a similar
conclusion about the LEG data.
If the HEG absorption line
were due to blueshifted absorption from lighter elements, one
would expect absorption features from other ionic species, particularly
Fe. If the line is due to a $K\alpha$  transition from \fexxv or \fexxvip,
the observed redshift then corresponds to a recession velocity
of $20750 ^{+585}_{-810} \ \rm km \ s^{-1}$ or
$32130 ^{+560}_{-780} \ \rm km \ s^{-1}$ respectively
(see Yaqoob \etal 2003b and references therein for the sources
of atomic data that we use).
At the very least, $v/c \sim 3\%$, or $\sim 9380 \ \rm km \ s^{-1}$
(corresponding to Fe~{\sc i}).

At face value, one could interpret the absorption-line
redshift of $\sim 2-3 \times 10^{4} \rm \ km \ s^{-1}$ as
inflow (e.g. onto the putative central black hole).
To derive a recession velocity one would have to account
for gravitational redshifts (at the
unknown radial distances).
Evidence for inflow is very rare, and in the X-ray band,
only one other case has been reported (in NGC 3516, Nandra \etal 1999).
In that case inflow was not the only interpretation
(e.g. see Ruszkowski \& Fabian 2000) and neither
is it for the present case of \srcp.

Another interpretation is that the absorption line is due to
highly-ionized
material that is crossing the line-of-sight obliquely and may,
for example, have have been ejected from the accretion disk
(a similar scenario for line {\it emission} was considered for
Mkn~766 by Pounds \etal 2003a and Turner, Kraemer, and Reeves 2004).
If the ejected blob still has the Keplerian motion of the disk
then from the redshift and upper limit on the absorption
line width, one can derive  a lower limit on the radial distance
from the black hole since, in a given observation time,
the blob cannot travel too far or else the
absorption line would be too broad. This gives
a constraint on the minimum radial distance of the absorption
site from the central black hole, as a function of the disk inclination angle.
This scenario
predicts that the absorption line energy and EW should be
significantly variable.

Yet another possibility
is that the redshift of the absorption line is
gravitational in origin. The absorber could of course be
outflowing and yet still give an overall redshift, if it
is close enough to the putative central black hole.
From pure gravitational redshifts alone, one can place a limit
on the allowed range of distances, $r$, from the black hole that
can be occupied by the absorbing gas. To first order, from a given
line width $\Delta E$, $\Delta r \sim r (r/r_{g}) (\Delta E/E_{0})$,
where $E_{0}$ is the rest energy of the line. For $r = 10r_{g}$,
$\Delta E \sim 50$~eV, and $E_{0} \sim 7$~keV, we have $\Delta r \sim r_{g}$.
Obviously, the actual allowed range of radii
required to keep the absorption line narrow depends on the details
of the geometry of the X-ray source and of the kinematics
of the absorber (whether inflow or outflow). The
detailed kinematics of the flow, its physical state (e.g. ionization,
density, temperature), and its proximity to the putative black
hole will all affect the line width. Nevertheless, $\Delta r$
is rather small. However, we note that the quasar model
of Elvis (2000) involves a cylindrical outflow
(that could intercept the X-ray continuum over a narrow range of radii).
Reports of outflows in AGN are now common, and a few are claimed
to originate within tens to hundreds of gravitational radii
of the central black hole (e.g. Pounds \etal 2003b, 2003c).
However, it has been claimed that many of these outflows may
not exist if absorption features that are really local to our
Galaxy have been misidentified as being at the redshift of the
AGN (McKernan, Yaqoob, \& Reynolds 2005).

\begin{table}[hbt]
\caption{Gaussian Line Fits to \asca SIS Data for E~1821$+$643}
\label{tab:npagetab}
\begin{tabular}{@{}l@{\extracolsep{\fill}}r}
& \\
\hline
Parameter & Measurement \\
\hline
& \\
$\chi^{2}$ & 168.6 \\
Degrees of freedom & 266 \\

Emission Line Rest Energy (keV) & $6.62^{+0.12}_{-0.12}$ \\
Emission Line Intensity & $3.4^{+1.9}_{-2.1}$ \\
($\rm 10^{-5} \ photons \ cm^{-2} \ s^{-1}$) & \\

Emission Line Equivalent Width & $140^{+78}_{-86}$ \\
(eV) & \\
$\sigma$, Gaussian Emission Line Width & $0.14^{+0.16}_{-0.14}$ \\
(keV) & \\
FWHM of Emission Line ($\rm km \ s^{-1}$) & $14935^{+17660}_{-14935}$ \\
Absorption Line Equivalent Width, & $4^{+78}_{-4}$ \\
EW (eV) & \\
Power-Law Photon Index, $\Gamma$ & $1.80^{+0.03}_{-0.02}$ \\
2--10 keV Observed Flux & 1.6 \\
($10^{-11} \rm \ ergs \ cm^{-2} \ s^{-1}$) & \\
2--10 keV Luminosity, & 4.3 \\
Quasar Frame ($10^{45} \rm \ ergs \ s^{-1}$) & \\
\hline
\end{tabular}
{\small
Simple power-law model plus Gaussian emission line
fitted to the {\it ASCA} SIS data
(see \figascalegratsp, \figascalegprofilesp, \figmultimissioncontp,
and \S\ref{ascafitting} for details).
All parameters (except 2--10~keV flux) refer to the quasar frame
($z=0.297$).
Errors are 90\% confidence for one parameter ($\Delta C = 2.706$).
The EW of a Gaussian absorption line added to the above model
is also given, where the center energy and width of the
absorption line are fixed at the respective values obtained from
the HEG data (see \tablegaussfitsp).
Velocities have been rounded to the nearest 5~$\rm km \ s^{-1}$.
Intrinsic luminosity, in the 2--10~keV band in the quasar frame,
was calculated using $H_{0} = 70 \rm \ km \ s^{-1} \ Mpc^{-1}$ and $\Lambda = 0.7
$.}
\end{table}

\section{CONCLUSIONS}
\label{conclusions}

During a \chandra
\hetg observation of \src we detected an absorption line, at a significance level of $2-3\sigma$, at $\sim 6.2$~keV in the
quasar frame. Whether or not this absorption line is
accounted for in modeling the Fe-K {\it emission line},
directly affects the inferred width of the emission line and
its peak energy.
We have also found that the Fe-K emission-line spectra of \src during
observations with the \chandra \hetgp, \chandra \letgp, and
\asca appear at first
to be different to each other. The first two observations were
separated by $\sim 3$~weeks and the \asca observation was
made nearly eight years earlier.
The 2--10~keV luminosity
of \src during the \asca and \hetg observations was only $\sim 60\%$,
and $\sim 20\%$ higher, respectively, than during the \letg
observation. The Fe-K line profile during the \asca
observation was peaked at $\sim 6.6$~keV, during
the \hetg observation it was strongly peaked at $\sim 6.4$~keV,
and during the \letg observation the line profile was double-peaked,
at $\sim 6.4$~keV, and $\sim 6.9$~keV (all are quasar-frame
energies).
These
properties of the Fe-K line region are captured in
\figascalegratsp, \figascalegprofilesp, and \figmultimissioncontp.
However, due to limited signal-to-noise, a non-varying
Fe-K complex cannot be ruled at 99\% confidence.
It is possible that $K\alpha$ emission from
from Fe~{\sc i--xvii} or so,
He-like Fe, and \feklya is present with similar intensities in all
three data sets. However, it will be extremely important to
test for variability in the Fe-K emission complex with  future
missions.

Clearly, the structure of the line-emitter
must be complex (for example, there may be more than one emitter),
in order to produce this range in Fe ionization
states {\it simultaneously}
(e.g. emission lines from Fe~{\sc i-xvii}
and \fexxvi are detected with a high significance
in the \letg data).
However, the location of the line-emitting region
or regions is not known.
Certainly, in the \hetg data the Fe-K line
at $\sim 6.4$~keV could be broad,
indicating
an origin close to the putative black hole, such as an accretion
disk. The spectral resolution of the \asca and LEG data
is insufficient to ascertain whether the emission lines
during these observations were narrow or broad, although
it appears that during the \letg observation the lines were
not as broad as the line observed during the \hetg observation.

The putative absorption line
detected by the high energy grating (HEG) on the \chandra \hetg
is not unequivocally ruled out
by any of the data sets.
We argue that if the absorption
line is real, it is most likely due
to redshifted resonance absorption in either He-like or
H-like Fe. The origin of the redshift
(equivalent to $v/c \sim 0.066-0.106$), could be absorption
in highly ionized matter that is either inflowing
or crossing the line-of-sight obliquely. In the latter case
the material could be due to ejecta from the accretion disk.
A more controversial interpretation is that the absorption
line is due to an {\it outflow} that is so close to the
black hole ($<15r_{g})$, that the absorption line is gravitationally redshifted.
If this is true then we would have an important
new diagnostic for studying strong gravity, complementing
studies using the \fekalfa emission lines.

{\it AstroE-2}, with a factor of $\sim 6$ greater
effective area than the \chandra HEG at the relevant energies,
and a spectral resolution of $\sim 6$~eV, will
help to constrain the origin of the
Fe-K emission features in \srcp, and provide a more sensitive search
for absorption lines of the kind found in the HEG data.

TY gratefully acknowledges support from
NASA grants NNG04GB78A and NAG5-10769, and
AR4-5009X, the latter issued by the
Chandra X--ray Observatory Center, operated by the SAO for and on behalf of
NASA under contract NAS8--39073.
This research
made use of the HEASARC online data archive services, supported
by NASA/GSFC and also
of the NASA/IPAC Extragalactic Database (NED) which is operated by
the Jet Propulsion Laboratory, California Institute of Technology,
under contract with NASA.
The authors are grateful to the \chandra
instrument and operations teams, and to
Urmila Padmanabhan for help with some of the analysis.

\section*{\bf FULL FIGURE CAPTIONS}

\par\noindent
{\bf Figure 5} ({\it Full Caption}) \\
(a) Contours corresponding to joint two-parameter
confidence levels of 68\%, 90\%, and 99\% for the intensity
and center energy (observed frame) of the broad \fekalfa line in \src when
it is modeled with a simple Gaussian.
Black contours correspond to the case
when an inverted Gaussian is included to model
the absorption line in the HEG data (see \figabscontoursp)
and red contours to the case when
the absorption line is omitted from the model and these
contours are entirely consistent with Fig.~4 of Fang \etal (2002).
The red contours are shown again in \figmultimissioncont of the
present paper, in that case plotted against {\it quasar-frame}
energy, and these are directly comparable with
Fig.~4 of Fang \etal (2002).
Excluding the absorption line results in a slightly higher
peak energy for the emission line.
The vertical dotted line corresponds to 6.4~keV in the
quasar frame.
(b)
As (a) for the equivalent width
(EW) and FWHM of the broad \fekalfa line.
The dashed line corresponds to the HEG FWHM
spectral resolution at the observed peak energy of the broad line,
and shows that the \fekalfa line is resolved by the HEG.
The absorption line is included in the model (see \figabscontoursp).
See \S\ref{gaussianfitting} and \tablegaussfits for details.
\\
\\
{\bf Figure 7} ({\it Full Caption}) \\
Comparison of the Fe-K complex in \src between non-contemporaneous
\chandra HEG, \chandra LEG, and \asca (SIS) data (see \S\ref{otherdata}).
Shown are the ratios of data to a simple power-law model, fitted to each
data set in the 1--9~keV observed energy range
(excluding the 2--2.5~keV interval for the HEG and LEG data).
It can be seen that the line profile in each case appears to
be different, with the LEG profile showing
two peaks (but see \S\ref{legfitting} and \S\ref{ascafitting}
for a quantitative assessment of variability).
The vertical dotted lines correspond (from left to right) to the
energies of the following transitions: Fe~{\sc i}~K$\alpha$, \fexxv
\resonetwo
resonance, and \feklya (that
correspond to rest-frame energies of 6.400, 6.700, and 6.966~keV
respectively).
It can be seen that the HEG profile peaks near
Fe~{\sc i}~K$\alpha$, the LEG peaks are near
Fe~{\sc i}~K$\alpha$ and \feklyap, whilst the \asca peak is near
the \fexxv triplet resonance line energy.
\\
\\
{\bf Figure 9} ({\it Full Caption}) \\
Direct comparison of
joint, two-parameter, 68\%, and 99\% confidence contours of emission-line
intensity, $I$, versus line energy between
non-contemporaneous
\chandra HEG, \chandra LEG, and \asca (SIS) data (see \S\ref{otherdata}).
For \ascap, 90\% confidence contours are also shown.
In each case the data were modeled with a simple power law and
a Gaussian emission line (two in the case of the LEG in order to
model the two peaks - see \figascalegratsp).
The black HEG contours were obtained when an inverted gaussian
was included to model the absorption feature (the same contours
are shown in \figlinecontoursp). The red HEG contours were
obtained with no absorption feature included (as were the LEG and \asca contours).
The red HEG contours are consistent with, and directly comparable with
those shown in Fig.~4 of Fang \etal (2002).
The vertical dotted lines correspond (from left to right) to the
energies of the following transitions: Fe~{\sc i}~K$\alpha$, and
\fexxv \resonetwo
resonance, and \feklya
(that correspond to rest-frame energies of 6.400, 6.700, and 6.966~keV
respectively). At 99\% confidence the evidence for variability in the
{\it total} line intensity between LEG and HEG (i.e. for LEG this means the sum of the two
lines) is marginal, and the \asca 99\% contours are large enough to
overlap all the other contours.
The \asca emission line can also be modeled as two discrete,
unresolved lines at 6.4~keV and 6.7~keV, and the 99\% two-parameter ranges
on the line intensities for this scenario are shown as solid vertical bars
at those energies (also see \figascaivsip).

\end{document}